\documentclass[a4paper,twoside,twocolumn,spanish,english]{revtex4-1}
\usepackage[T1]{fontenc}
\usepackage[latin9]{inputenc}
\usepackage{textcomp}
\usepackage{amsmath}
\usepackage{amssymb}
\usepackage{graphicx}
\usepackage{esint}

\makeatletter

\usepackage{braket}

\usepackage{babel}

\usepackage{babel}

\makeatother

\usepackage{babel}
\addto\shorthandsspanish{\spanishdeactivate{~<>}}

\begin{document}
\selectlanguage{spanish}%
\global\long\def\abs#1{\left|#1\right|}
\global\long\def\ket#1{\left|#1\right\rangle }
\global\long\def\bra#1{\left\langle #1\right|}
\global\long\def\half{\frac{1}{2}}
\global\long\def\partder#1#2{\frac{\partial#1}{\partial#2}}
\global\long\def\comm#1#2{\left[#1,#2\right]}
\global\long\def\vp{\vec{p}}
\global\long\def\vpp{\vec{p}'}
\global\long\def\dt#1{\delta^{(3)}(#1)}
\global\long\def\Tr#1{\textrm{Tr}\left\{  #1\right\}  }
\global\long\def\Real#1{\mathrm{Re}\left\{  #1 \right\}  }
\global\long\def\braket#1{\langle#1\rangle}
\global\long\def\escp#1#2{\left\langle #1|#2\right\rangle }
\global\long\def\elmma#1#2#3{\langle#1\mid#2\mid#3\rangle}
\global\long\def\ketbra#1#2{|#1\rangle\langle#2|}

\selectlanguage{english}%

\title{Mimicking spatial localization in dynamical random environment}

\author{I. Márquez-Martín$^{1}$}

\author{G. Di Molfetta$^{1,2}$}

\author{A. Pérez$\text{\textonesuperior}$}

\affiliation{$^{1}$Departamento de Física Teórica and IFIC, Universidad de Valencia-CSIC,
Dr. Moliner 50, 46100-Burjassot, Spain\\
 $^{2}$Aix-Marseille Universite, CNRS, Laboratoire d'Informatique
Fondamentale, Marseille, France}
\begin{abstract}
We study the role played by noise on the QW introduced in \cite{Ahlbrecht2012},
a 1D model that is inspired by a two particle interacting QW. The
noise is introduced by a random change in the value of the phase during
the evolution, from a constant probability distribution within a given
interval. The consequences of introducing such kind of noise depend
on both the center value and the width of that interval: a wider interval
manifests as a higher level of noise. For some range of parameters,
one obtains a quasi-localized state, with a diffusive speed that can
be controlled by varying the parameters of the noise. The existence
of this (approximately) localized state for such kind of \textit{time-dependent}
noise is, to the best of our knowledge, totally new, since localization
(i.e., Anderson localization) is linked in the literature to a \textit{spatial
random} noise. 
\end{abstract}
\maketitle

\section{Introduction}

Quantum walks (QWs) are formal quantum analogues of classical random
walks. QWs were first considered by Gr{ö}ssing and Zeilinger \cite{grossing1988quantum}
in 1988, as simple quantum cellular automata in the one-particle sector.
They were popularized in the physics community in 1993, by Y. Aharonov
\cite{Y.AharonovL.Davidovich1993}, who started studying them systematically
and described some of their main properties. The formalism was then
developed a fews years later by D.A. Meyer \cite{meyer1996quantum},
in the context of quantum information. From a physical perspective,
QWs describe situations where a quantum particle is taking steps on
a lattice conditioned on its internal state, typically a (pseudo)
spin one half system. The particle dynamically explores a large Hilbert
space associated with the positions of the lattice, and thus allows
to simulate a wide range of transport phenomena \cite{Kempe2003}.
With QWs, the transport is driven by an external discrete unitary
operation, which sets it apart from other lattice quantum simulation
concepts where transport typically rests on tunneling between adjacent
sites \cite{bloch2012quantum}: all dynamical processes are discrete
in space and time. As models of coherent quantum transport, QWs are
interesting both for fundamental quantum physics and for applications.
An important field of applications is quantum algorithmic \cite{ambainis2003quantum}.
QWs were first conceived as the natural tool to explore graphs, for
example for efficient data searching (see e.g. \cite{magniez2011search}).
QWs are also useful in condensed matter applications and topological
phases \cite{kitagawa2012observation}. A totally new emergent point
of view for QWs concerns quantum simulation of gauge fields and high-energy
physical laws \cite{PhysRevA.94.012335,genske2013electric}. It is
important to note that QWs can be realized experimentally with a wide
range of physical objects and setups, for example as transport of
photons in optical networks or optical fibers \cite{schreiber20122d},
or of atoms in optical lattices \cite{cote2006quantum}. Although
most of these families of QWs represent and model the one-particle
dynamics, implementation of the two-particle sector has been investigated
and studied by several authors \cite{Meyer1997,sansoni2012two,meyer1996quantum}. In
particular a two-particle QW, modeling an atom-atom binding in free
space, has been already introduced by \cite{Ahlbrecht2012} in a very
general framework. The latter model can be reduced to an effective
one-particle QW that contains some of the properties of the two-particle
dynamics, in the sense that it reproduces the same spectral properties
as the original model.

The main aim of this paper, inspired by the latter model, is to investigate
the role of dynamical noise. As we show, adding such kind of noise
produces a shape on the probability distribution that is characteristic
of random spatial noise. Moreover, by varying the strength of the
noise, one can obtain a quasi-localized distribution, with a diffusion
constant that can be made very small. The existence of this quasi-localized
state for such kind of time-dependent noise is, to the best of our
knowledge, totally new, since localization (i.e., Anderson localization)
has been observed and studied in a spatial random environment. Anderson
localization has also been studied in the context of quantum walks,
see for example \cite{Joye2010,Albrecht2011}. We support all results
by direct numerical simulations of the average dynamics, and by an
analytical calculation of the diffusion constant

The paper is organized as follows. Section II is devoted to the definition
of the model, whose main properties are discussed in Section III.
In Section IV we study numerically the behavior of this QW in a dynamical
random environment, and we show the existence of quasi-localized states. Section
\ref{sec:Discussion} summarizes our conclusions. The details about
the analytical calculation of the diffusive constant have been relegated
to the Appendix.

\section{Quantum walk on a line }

The dynamics of the QW takes place on the Hilbert space $\mathcal{{H}=\mathcal{{H}}}_{p}\otimes\mathcal{{H}}_{c}$,
where $\mathcal{{H}}_{p}$ is defined over $\mathbb{Z}^d$, and $\mathcal{{H}}_{c}$ corresponds
to the internal degree of freedom, which is usually referred to as
the coin space, with basis $\{\ket{c_{i}}\}_{i\in\mathcal{{K}}=[1,..,k]}$.
In the simplest case, it could describe a particle moving in a lattice
of dimension $d$, and the coin could correspond to the spin. The
unitary matrix which describes the evolution is $U=S(\mathbb{{I}}\otimes C)$.
The operator $S$ acts on the Hilbert space $\mathcal{{H}}$ as a
conditional position shift operator, with the coin acting as a control
qubit \cite{Nielsen2011}, which means that it will shift the position
of the particle to the right if the coin points up, and to the left,
if the coin points down. Therefore, choosing $d=1$:

\begin{equation}
S=\underset{i\in\mathcal{{Z}},}{\sum}\ket{i+1}\bra i\otimes\ket{\uparrow}\bra{\uparrow}+\ket{i-1}\bra i\otimes\ket{\downarrow}\bra{\downarrow}.\label{eq:shift}
\end{equation}

$\mathbb{{I}}\otimes C$ acts nontrivially only on the coin space
$\mathcal{{H}}_{c}$. The Hilbert space $\mathcal{{H}}$ is spanned
by the orthonormal basis $\{\Ket{i}\otimes\ket c:i\in\mathbb{{Z}},c\in{\uparrow,\downarrow}\}$.
$C$ is chosen as an element of $SU(2)$. The usual choice for this
matrix is given by the so called Hadamard coin:

\begin{equation}
C_{H}=\frac{1}{\sqrt{2}}\left(\begin{array}{cc}
1 & 1\\
1 & -1
\end{array}\right).
\end{equation}

After $t$ steps taken by the walker, the state becomes $\ket{\Psi_{t}}=U^{t}\Ket{\Psi_{0}}$.
The probability distribution at time step $t$ can be written as
\begin{equation}
P(x,t)=||\escp x{\Psi_{t}}||^{2}.\label{eq:Probdistribution}
\end{equation}

We now introduce a family of QWs defined by the coin operator
\begin{equation}
C_{g}=\frac{1}{2\gamma-1}\left(\begin{array}{cc}
\gamma & \sqrt{2}(\gamma-1)\\
\sqrt{2}(\gamma-1)\gamma & \gamma
\end{array}\right).\label{eq:Cgamma}
\end{equation}
This operator has been discussed in \cite{Ahlbrecht2012} as a way
to obtain a one-particle QW using the dispersion relation
corresponding to the relative motion of an interacting two-particle
QW, if one adopts a simple phase for the value of $\gamma$, i.e.
$\gamma=e^{ig}$ with $g$ a real parameter that controls the strength
of the coupling. In what follows, we will study the properties of
this model, regardless of its original motivation. We will show that
this one-particle QW possesses interesting properties on its own,
specially when dynamical noise is included.

\section{Noiseless case}

\begin{figure}
\includegraphics[width=1\columnwidth]{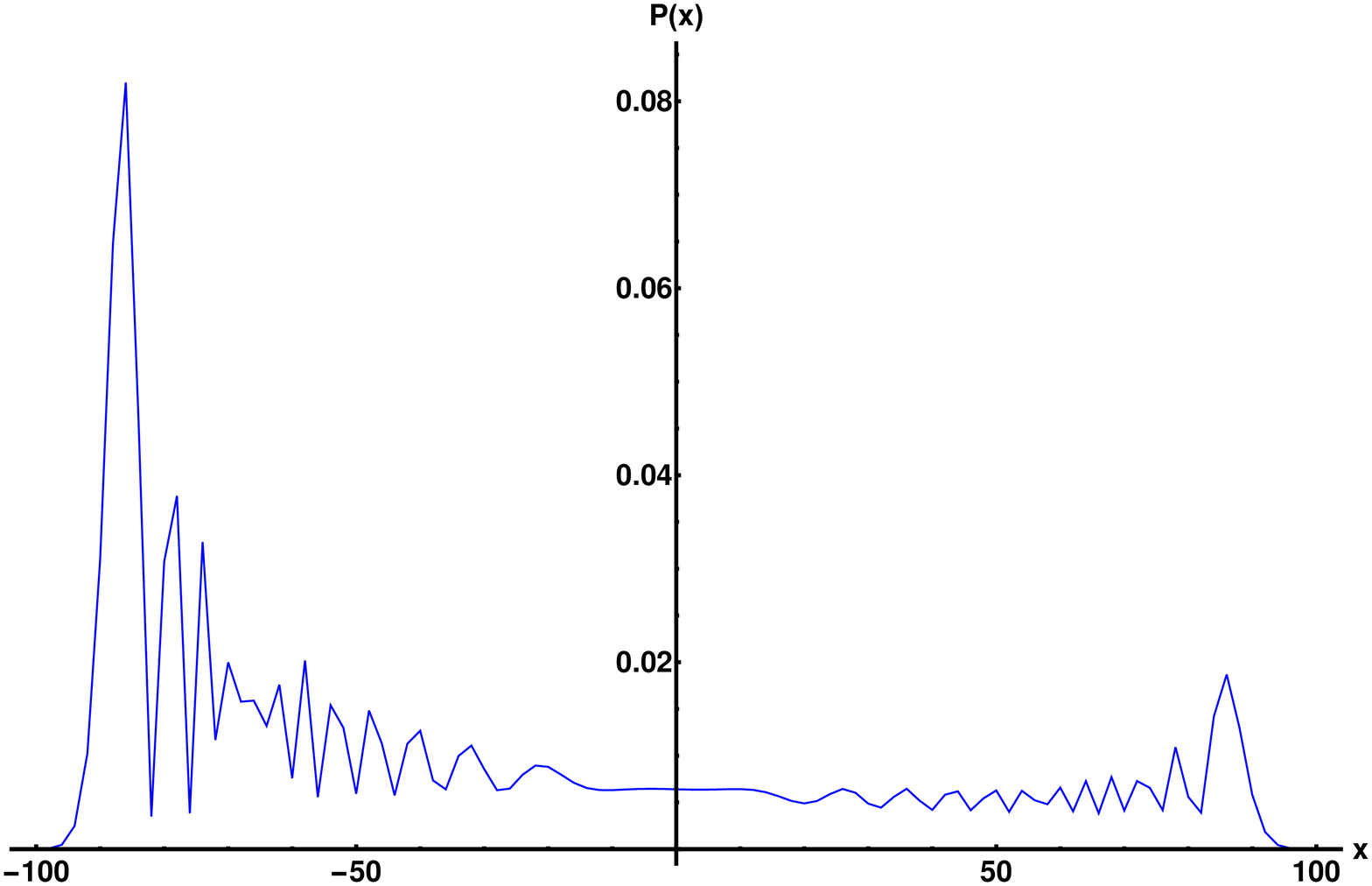}
\includegraphics[width=1\columnwidth]{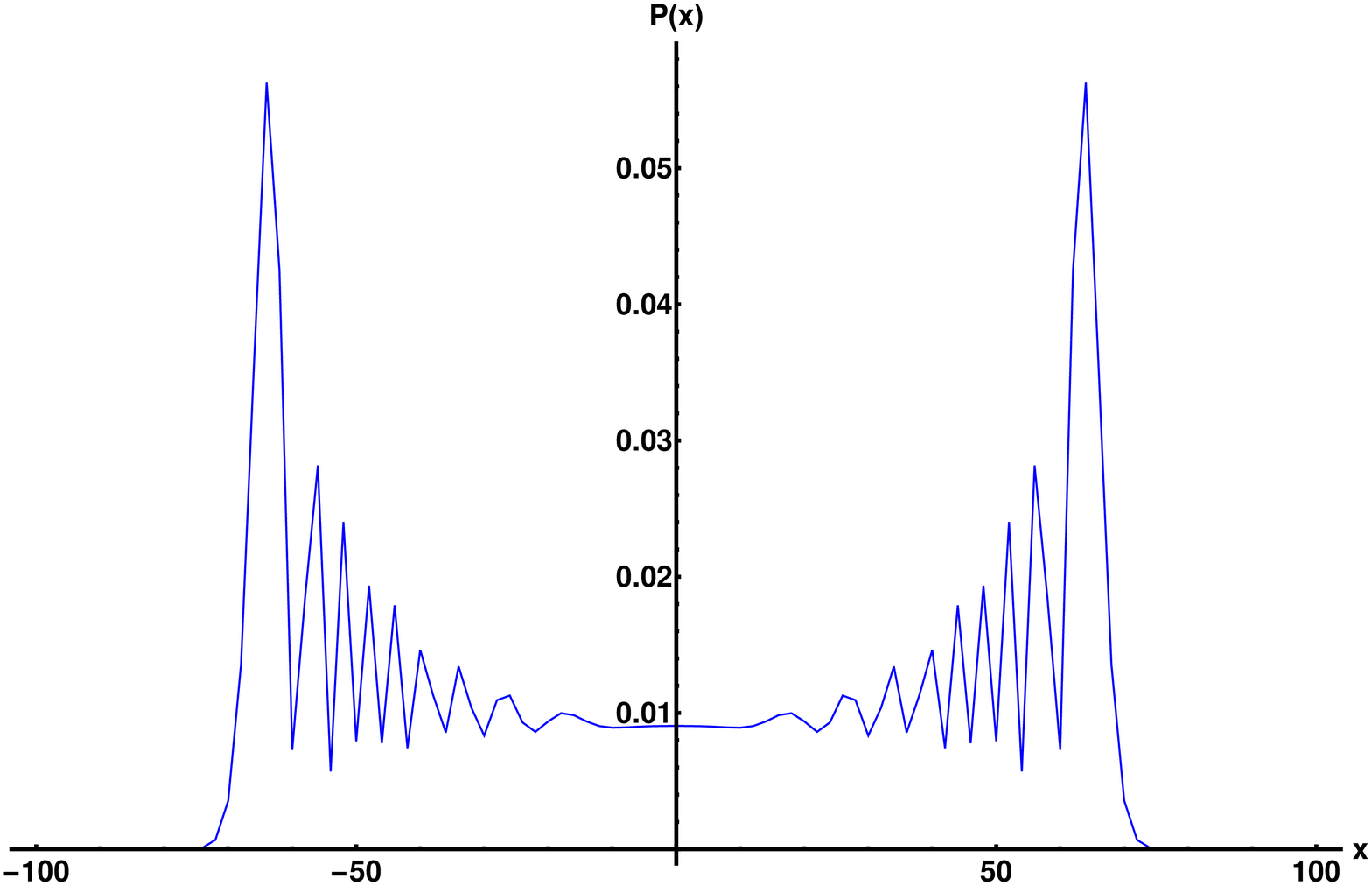}
\caption{(Color online) Probability distribution after $t=200$ time steps
for the QW, where only even sites are plotted (the probability vanishes
at odd sites, if $t$ is even). The initial state is localized at
the origin, with a coin state $\frac{1}{\sqrt{2}}(1,i)^{T}$. Upper
panel: QW with $g=\pi/2$. Lower panel: $g=\pi$.}
\label{fig:Pxtnonoise}
\end{figure}

Most properties of the QW are better analyzed by switching to the
quasi-momentum space \cite{Ambainis2001}. We introduce the basis
of states $\{\ket k,k\in[-\pi,\pi[\}$ defined by 
\begin{equation}
\ket k=\sqrt{\frac{1}{2\pi}}\sum\limits _{x=-\infty}^{\infty}e^{ikx}\ket x.\label{eq:pstates}
\end{equation}
The unitary operator that governs the QW can be written, in quasi-momentum
space, as 
\begin{align}
U(k,g) & =\left(\begin{array}{cc}
e^{-ik} & 0\\
0 & e^{ik}
\end{array}\right)C_{g}\nonumber \\
= & \frac{1}{2e^{ig}-1}\left(\begin{array}{cc}
e^{i(g-k)} & \sqrt{2}e^{-ik}\left(e^{ig}-1\right)\\
\sqrt{2}e^{i(g+k)}\left(e^{ig}-1\right) & e^{i(g+k)}
\end{array}\right).
\end{align}
Fig. \ref{fig:Pxtnonoise} shows the probability distribution for
two values of the parameter $g$. As one observes, the probability
resembles in both cases the one of a typical QW, although it can be
asymmetric, depending on the value of $g$ and on the initial coin
state.

\section{Dynamical random noise\label{sec:random_noise}}

\begin{figure}
\includegraphics[width=1\columnwidth]{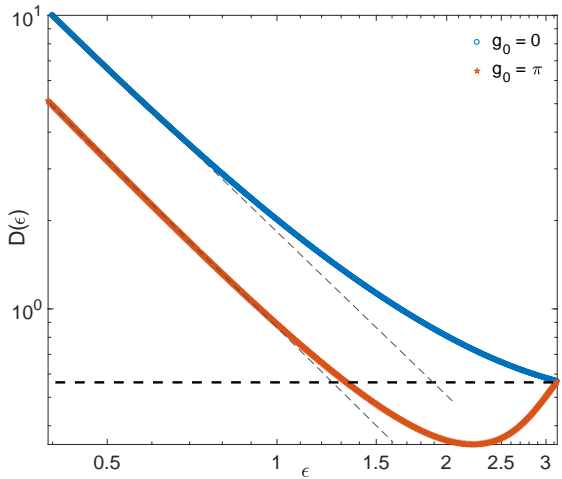}
\caption{(Color online) Diffusion constant, Eq. (\ref{eq:Depsilon}) for two
values of $g_{0}$: $g_{0}=0$ (upper curve), and $g_{0}=\pi$ (lower
curve), as a function of the parameter $\varepsilon$.}
\label{fig:diffwithepsilon}
\end{figure}

The main goal in this paper is the introduction of dynamical noise
for the model described in the previous section. In order to see how
the system responds to this kind of noise, we numerically simulate
its evolution using the same equations as in the noiseless case, but
randomly change the parameter $g$ at every step, from a uniform distribution, 
within the interval $[g_{0}+\text{\ensuremath{\epsilon}},g_{0}-\epsilon${]},
where $g_{0}$ is the center of the distribution. The entire procedure is
repeated over a number of $N_{s}$ iterations, and the result is averaged
over these samplings. As we discuss in the Appendix, this procedure
is equivalent to a description in terms of a Lindblad operator acting
on the density matrix that describes the state of the system. Using
this, one can compute the asymptotic value of the diffusion constant,
Eq. (\ref{eq:Depsilon}). The results are plotted on Fig. (\ref{fig:diffwithepsilon})
for two values of $g_{0}$, as a function of $\varepsilon$. As can
be observed from the plots, $D(\varepsilon)$ rapidly falls 
for not too large values of $\varepsilon$. With $g_{0}=0$, the curve
deviates from the initial trend as $\varepsilon$ approaches $\pi/4$.
The behavior for $g_{0}=\pi$ is different: $D(\varepsilon)$ has
a minimum value at $\varepsilon\simeq2.23$. Both curves have the
same value when $\varepsilon=\pi$, given by $D(\pi)=1-\sqrt{3}/4$. 

\begin{figure}
\includegraphics[width=1\columnwidth]{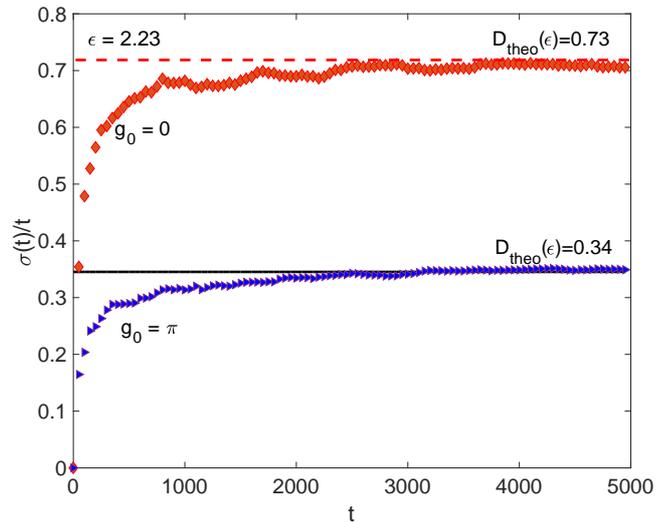}
\caption{(Color online) Diffusion constant for two
values of $g_{0}$: $g_{0}=0$ (upper panel), and $g_{0}=\pi$ (lower
panel), as a function of time, as obtained from the numerical simulation of the
noisy QW. The horizontal lines show the asymptotical analytical result Eq. (\ref{eq:Depsilon}).}
\label{fig:comparesigmaovert}
\end{figure}

The approach to the asymptotic regime can be observed from Fig. (\ref{fig:comparesigmaovert}).
In this figure, we plot the evolution with time of the noisy QW, as
numerically obtained following the procedure explained at the beginning
of this Section, in comparison with our analytical result Eq. (\ref{eq:Depsilon}),
for the value $\varepsilon=2.23$. We observe a good agreement between
both approaches.

The strong decrease of the diffusion constant observed in Fig. (\ref{fig:diffwithepsilon})
can be used to control the spreading of the QW, so that the wavepacket
expands at very low speed, as compared with the usual noiseless case.
Such decrease can, in principle, be further reduced by the use of
a non-localized initial state, since this amounts to including the
initial shape $f(k)$ of the wavepacket in the integral Eq. (\ref{eq:x2Gamma}). 

\begin{figure}
\includegraphics[width=1\columnwidth]{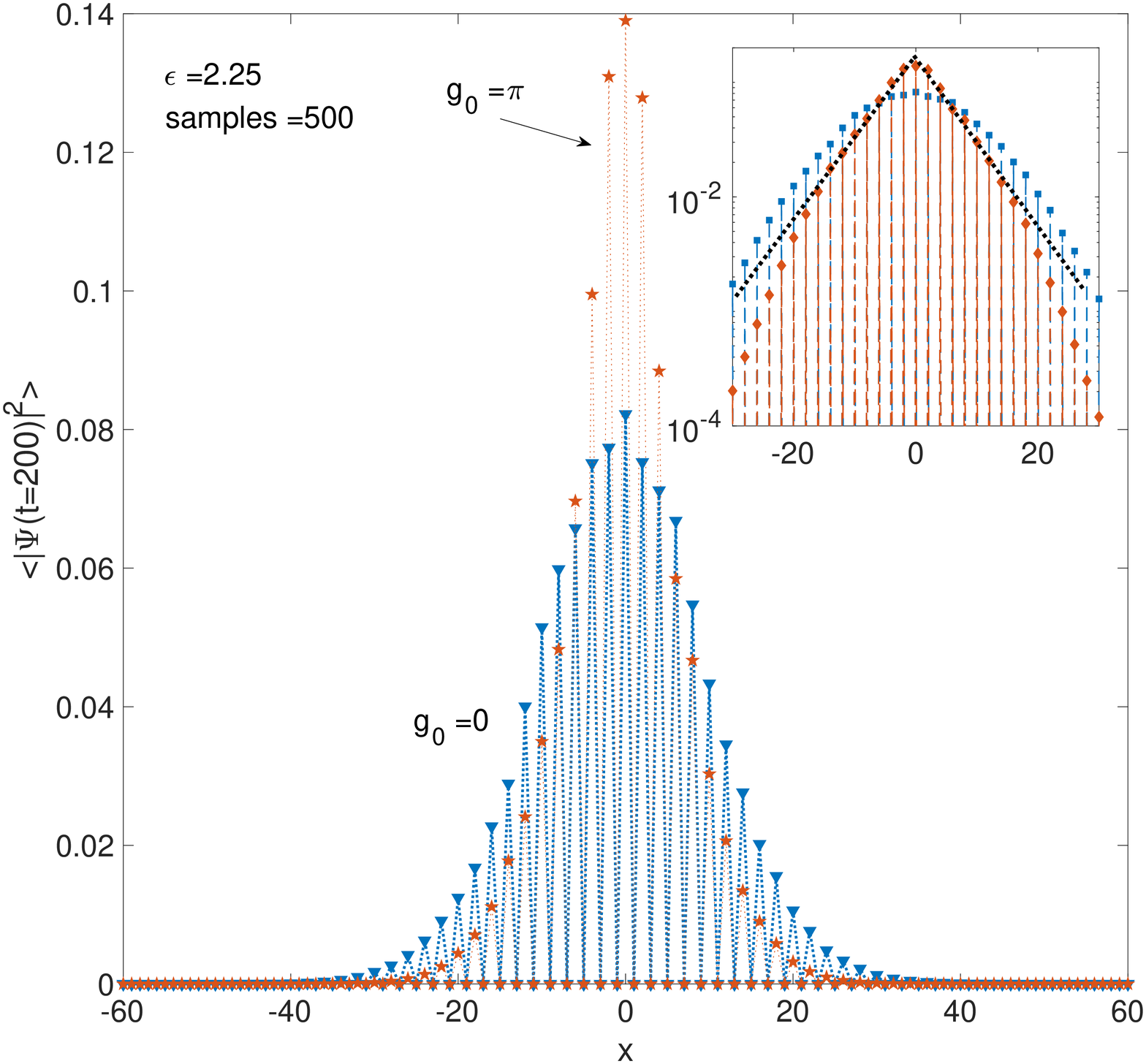}
\caption{(Color online) Evolved probability distribution, in lin-lin plot and log-ling plot (the inset), for two values of
$g_{0}$: $g_{0}=0$ (blue triangles), and $g_{0}=\pi$ (red stars)
at time step $t=200$, for an initially localized state at the origin. }
\label{fig:Profile}
\end{figure}

Even more interesting is the study of the probability distribution,
Eq. (\ref{eq:Probdistribution}). We can observe how the probability
concentrates around $x=0$ (see Fig. \ref{fig:Profile}) . Differently from the case $g_{0}=0$, for which we obtain the expected Gaussian solution, in the case $g_{0}=\pi$ we observe an exponential shape around zero. In fact, this evolution reminds us the phenomenon of Anderson localization. The result is intriguing due
to the fact that dynamical noise is expected to transform the system
into a classical random walk, but in this case we see some type of
dynamical localization. Anderson localization has extensively been
studied \cite{Evers2008,Lagendijk2009,Ahlbrecht2011b,Joye2011} as
the result of introducing a static disorder. It has also been studied
in the context of quantum walks, see for example \cite{Joye2010,Albrecht2011}.
What we obtained here, however, is a similar result, but produced
by a time-dependent noise. This is the main result of our paper.

\section{Conclusions}

\label{sec:Discussion}

In this paper, we investigated the role played by noise on the QW
introduced in \cite{Ahlbrecht2012}, a 1D model that is inspired by
a two particle interacting QW. The noise is introduced by a random
change in the value of $g$ during the evolution, from a constant
probability distribution within a given interval. The consequences
of introducing such kind of noise depend on both the center value
and the width of that interval: a wider interval manifests as a higher
level of noise. We observe that, by appropriately choosing the
level of noise, one obtains a quasi-localized state, which a spreading
rate can be made very small. The existence of this quasi-localized
state for such kind of \textit{time-dependent} noise is, to the best
of our knowledge, totally new, since localization (i.e., Anderson
localization) is linked in the literature to a \textit{spatial random}
noise.

\section{Acknowledgements}

This work has been supported by the Spanish Ministerio de Educación
e Innovación, MICIN-FEDER project FPA2014-54459-P, SEV-2014-0398 and
Generalitat Valenciana Grant GVPROMETEOII2014-087.

\appendix

\section{Appendix: Diffusion constant}

In this Section we analyze the long term behavior of the diffusion
constant. Our calculation closely follows the formalism presented
in \cite{Brun2003a,Annabestani2009}. However, the dynamics of the
decoherent QW will be represented by a $4\times4$ matrix acting on
four-vectors, instead of a superoperator. At a given time, the state
of the system is defined, in quasi-momentum space by the vector
\begin{equation}
R_{\alpha}(k,k',t)=Tr\{\sigma_{\alpha}\elmma k{\rho(t)}{k'}\},\label{eq:vectorR}
\end{equation}
where $\alpha=1,2,3,4$, $\rho(t)$ is the density operator at time
step $t\in\mathbb{N}$, and $\sigma_{0}$ is the 2-dimensional identity,
while $\sigma_{i}$, $i=1,2,3$ are the Pauli matrices. The trace
in the latter equation is performed on the spin space. The above equation
can be inverted to give
\begin{equation}
\elmma k{\rho(t)}{k'}=\frac{1}{2}\sum_{\alpha=0}^{3}R_{\alpha}(k,k',t)\sigma_{\alpha}.\label{eq:vectordecomrho}
\end{equation}

The probability distribution at time t can be obtained from
\begin{equation}
P(x,t)=\int_{-\pi}^{\pi}\frac{dk}{2\pi}\int_{-\pi}^{\pi}\frac{dk'}{2\pi}e^{ix(k-k')}R_{0}(k,k',t).
\end{equation}

For a given sequence of choices $\{g_{1},g_{2},\dots g_{t}\}$, the
time evolution of the density operator is obtained as
\begin{equation}
\elmma k{\rho(t+1)}{k'}=U(k,g_{1},\dots g_{t})\elmma k{\rho(t)}{k'}U^{\dagger}(k',g_{1},\dots g_{t}),\label{eq:evolseqg}
\end{equation}
where $U(k,g_{1},\dots g_{t})\equiv U(k,g_{t})\dots U(k,g_{1})$.
We now assume that the above sequence is randomly obtained from an
interval $\{g_{0}-\epsilon,g_{0}+\epsilon\}$ with an uniform probability
distribution $p(g)=\frac{1}{2\epsilon}$ if $|g-g_{0}|\le\epsilon$,
and $p(g)=0$ otherwise. Under these assumptions, it has been proven
\cite{DiMolfetta2016} that, in the limit $t\rightarrow\infty$, Eq.
(\ref{eq:evolseqg}) can be described by the action of a superoperator
$\mathcal{\hat{L}}_{k,k'}$ defined as 
\begin{equation}
\mathcal{L}_{k,k'}\elmma k{\rho(t)}{k'}\equiv\frac{1}{2\varepsilon}\int_{g_{0}-\varepsilon}^{g_{0}+\varepsilon}dg\,\,U(k,g)\elmma k{\rho(t)}{k'}U^{\dagger}(k',g),
\end{equation}
so that 
\begin{equation}
\elmma k{\rho(t+1)}{k'}=\mathcal{\hat{L}}_{k,k'}\elmma k{\rho(t)}{k'}.
\end{equation}

We can recast the latter equation under algebraic form (with vectors
and matrices) by the use of Eq. (\ref{eq:vectordecomrho}). Then the
vector defined in Eq. (\ref{eq:vectorR}) evolves according to
\begin{equation}
R_{\alpha}(k,k',t)=\sum_{\beta=0}^{3}[\mathcal{L}_{k,k\text{\textasciiacute}}]_{\alpha\beta}R_{\alpha}(k,k',t-1),
\end{equation}
where $\mathcal{L}_{k,k\text{\textasciiacute}}$ is a $4\times4$
matrix defined as
\begin{equation}
[\mathcal{L}_{k,k\text{\textasciiacute}}]_{\alpha\beta}=\frac{1}{2\varepsilon}\int_{g_{0}-\varepsilon}^{g_{0}+\varepsilon}dg\,\,Tr\{\sigma_{\alpha}U(k,g)\sigma_{\beta}U^{\dagger}(k',g)\}.
\end{equation}
with $\alpha,\beta=0,1,2,3$. We are interested in deriving an analytical
expression for the variance 
\begin{equation}
\left\langle x^{2}\right\rangle _{t}=\sum_{x}x^{2}P(x,t),
\end{equation}
for large values of the time step $t$. The initial state is assumed
to be localized at $x=0$, therefore 
\begin{equation}
\rho(0)=\ket 0\bra 0 \otimes \rho_{c},
\end{equation}
where $\rho_{c}$ is the initial coin state. Then 
\begin{equation}
R_{\alpha}(k,k',0)=Tr\{\sigma_{\alpha}\rho_{c}\}\equiv r_{\alpha}
\end{equation}
is independent of both $k$ and $k'$, with $r_{0}=1$. Following
similar steps as in \cite{Annabestani2009}, one arrives to
\begin{eqnarray}
\left\langle x^{2}\right\rangle _{t} & = & \int_{-\pi}^{\pi}\frac{dk}{2\pi}\sum\limits _{m=1}^{t}\sum\limits _{m'=1}^{m-1}[\mathcal{G}_{k}^{\dagger}\mathcal{L}_{k}^{m-m'-1}\mathcal{G}_{k}\mathcal{L}_{k}^{m'-1}\nonumber \\
 & + & \mathcal{G}_{k}\mathcal{L}_{k}^{m-m'-1}\mathcal{G}_{k}^{\dagger}\mathcal{L}_{k}^{m'-1}]_{0\beta}r_{\beta}\nonumber \\
 & + & \int_{-\pi}^{\pi}\frac{dk}{2\pi}\sum\limits _{m=1}^{t}\sum_{\beta=0}^{3}[\mathcal{J}_{k}\mathcal{L}_{k}^{m-1}]_{0\beta}r_{\beta},\label{eq:x2t}
\end{eqnarray}
with the convention of sum over repeated indices, and the following
definitions
\begin{eqnarray}
\mathcal{L}_{k} & = & \left.\mathcal{L}_{k,k\text{\textasciiacute}}\right|_{k'=k}\nonumber \\
\mathcal{G}_{k} & = & \left.\frac{\partial\mathcal{L}_{k,k\text{\textasciiacute}}}{\partial k}\right|_{k'=k}\nonumber \\
\mathcal{J}_{k} & = & \left.\frac{\partial^{2}\mathcal{L}_{k,k\text{\textasciiacute}}}{\partial k\partial k'}\right|_{k'=k}.
\end{eqnarray}
It can be shown that 
\begin{equation}
[\mathcal{L}_{k}]_{0\beta}=[\mathcal{L}_{k}]_{\beta0}=\delta_{\beta,0}.
\end{equation}
The same property holds for $\mathcal{J}_{k}$. Use was made of these
properties in deriving Eq. (\ref{eq:x2t}). Therefore, we can write
$\mathcal{L}_{k}$ in block-diagonal form
\begin{equation}
\mathcal{L}_{k}=\left(\begin{array}{cc}
1 & 0\\
0 & M_{k}
\end{array}\right),
\end{equation}
$M_{k}$ being a $3\times3$ matrix. This structure is obviously preserved
with time, i.e.
\begin{equation}
\mathcal{L}_{k}^{t}=\left(\begin{array}{cc}
1 & 0\\
0 & M_{k}^{t}
\end{array}\right).
\end{equation}
The explicit form of $\mathcal{L}_{k,k'}$ is cumbersome for general
values of $g_{0}$. We will here concentrate on two particular cases,
namely $g_{0}=0$ and $g_{0}=\pi$. In both cases, one can write
\begin{equation}
\mathcal{L}_{k,k'}=\left(\begin{array}{cccc}
\cos u & ic_{12}\sin u & 0 & -ic_{44}\sin u\\
0 & c_{22}\cos v & c_{23}\sin v & c_{24}\cos v\\
0 & c_{22}\sin v & -c_{23}\cos v & c_{24}\sin v\\
-i\sin u & -c_{24}\cos u & 0 & c_{44}\cos u
\end{array}\right)\label{eq:Lkkp}
\end{equation}
with the definitions $u=k-k'$ and $v=k+k'$. The coefficients $c_{ij}$
depend both on the value of $g_{0}$ and $\varepsilon$. We have dropped
this dependence for simplicity. For $g_{0}=0$ one has $c_{12}=\frac{\sqrt{2}\left(a-\frac{3\epsilon}{2}\right)}{3\epsilon}$,
$c_{22}=\frac{1}{4}+\frac{\sin\epsilon}{\epsilon}-\frac{a}{6\epsilon}$,
$c_{23}=\frac{\epsilon+4\sin(\epsilon)-6a}{4\epsilon}$, $c_{24}=\frac{2a-3\epsilon}{3\sqrt{2}\epsilon}$,
and $c_{44}=\frac{4a}{3\epsilon}-1$, while for $g_{0}=\pi$ we found
$c_{12}=\frac{-3\epsilon-2b+\pi}{3\sqrt{2}\epsilon}$, $c_{22}=-\frac{-3\epsilon+12\sin(\epsilon)-2b+\pi}{12\epsilon}$,
$c_{23}=\frac{\epsilon-4\sin(\epsilon)+6b-3\pi}{4\epsilon}$, $c_{24}=\frac{-3\epsilon-2b+\pi}{3\sqrt{2}\epsilon}$
, and $c_{44}=\frac{-3\epsilon-4b+2\pi}{3\epsilon}$. We have introduced
the notations $a=\arctan\left(3\tan\frac{\epsilon}{2}\right)$, $b=\arctan\left(3\cot\frac{\epsilon}{2}\right)$.
The last term in Eq. (\ref{eq:x2t}) can be easily evaluated, resulting
in 
\begin{equation}
\int_{-\pi}^{\pi}\frac{dk}{2\pi}\sum\limits _{m=1}^{t}\sum_{\beta=0}^{3}[\mathcal{J}_{k}\mathcal{L}_{k}^{m-1}]_{0\beta}r_{\beta}=t.
\end{equation}

Starting from Eq. (\ref{eq:Lkkp}), one obtains that $[\mathcal{G}_{k}^{\dagger}]_{0\beta}=-[\mathcal{G}_{k}]_{0\beta}$.
Therefore, the first two terms in Eq. (\ref{eq:x2t}) can be combined
to give
\begin{equation}
\int_{-\pi}^{\pi}\frac{dk}{2\pi}\sum\limits _{m=1}^{t}\sum\limits _{m'=1}^{m-1}[\mathcal{G}_{k}\mathcal{L}_{k}^{m-m'-1}(\mathcal{G}_{k}^{\dagger}-\mathcal{G}_{k})\mathcal{L}_{k}^{m'-1}]_{0\beta}r_{\beta}.\label{eq:x2t2first}
\end{equation}
Moreover, one can check that, $\forall p,q\in\mathbb{N}$
\begin{equation}
[\mathcal{G}_{k}\mathcal{L}_{k}^{p}(\mathcal{G}_{k}^{\dagger}-\mathcal{G}_{k})\mathcal{L}_{k}^{q}]_{0\beta}
\end{equation}
does not depend on the action of $\mathcal{L}_{k}^{q}$, which allows
us to drop this matrix off. Thus the integrand in Eq. (\ref{eq:x2t2first})
can be expressed in terms of the submatrix
\begin{equation}
\sum\limits _{m=1}^{t}\sum\limits _{m'=1}^{m-1}M_{k}^{m-m'-1}=(I_{3}-M_{k}^{-1})[tI_{3}+(I_{3}-M_{k}^{-1})(M_{k}^{t}-I_{3})]\label{eq:summmp}
\end{equation}
We have verified that all the eigenvalues of $M_{k}$ obey $0<\left|\lambda\right|<1$
, so that $M_{k}^{t}\to0$ in the long time limit. Therefore, the
last term in Eq. (\ref{eq:summmp}) will be time-independent at large
time steps. In what follows, we will omit that term, since it does
not affect our conclusions. The rest of the calculation is straightforward,
and we obtain
\begin{equation}
\left\langle x^{2}\right\rangle _{t\rightarrow\infty}=t[1-\int_{-\pi}^{\pi}\frac{dk}{\pi}\Gamma(k,\varepsilon)],\label{eq:x2Gamma}
\end{equation}
where
\begin{equation}
\Gamma(k,\varepsilon)=\frac{\alpha\cos2k+\beta}{\gamma\cos2k+\delta},
\end{equation}
with $\alpha=c_{12}c_{24}+c_{44}(c_{22}-c_{23})$, $\beta=c_{12}c_{23}c_{24}+c_{44}(c_{22}c_{23}-1)$,
$\gamma=c_{22}(c_{44}-1)-c_{23}c_{44}+c_{23}+c_{24}^{2}$, and $\delta=c_{22}c_{23}(c_{44}-1)+c_{23}c_{24}^{2}-c_{44}+1$.
The integral in Eq. (\ref{eq:x2Gamma}) can be expressed in terms
of standard integrals \cite{Gradshteyn}. By defining $r=\sqrt{1-\gamma^{2}/\delta^{2}}$,
we arrive at the result:
\begin{equation}
\left\langle x^{2}\right\rangle _{t\rightarrow\infty}=t[1-\frac{2\alpha(r-1)}{\gamma r}-\frac{2\beta}{\delta r}].
\end{equation}
From here, we define the diffusive constant 
\begin{equation}
D(\varepsilon)\equiv\frac{d\left\langle x^{2}\right\rangle _{t\rightarrow\infty}}{dt}=1-\frac{2\alpha(r-1)}{\gamma r}-\frac{2\beta}{\delta r}.\label{eq:Depsilon}
\end{equation}
 \bibliographystyle{apsrev4-1}
\bibliography{library}

\end{document}